\documentclass[preprint,nofootinbib]{revtex4-2}

\usepackage{graphicx}
\usepackage{dcolumn}
\usepackage{bm}

\begin{document}

\title{The nox-minima dome: a tangible alternative to flat star maps for astronomy education}

\author{A. Bessa}\email{andre.bessa@ufrn.br}
\affiliation{%
Escola de Ciências e Tecnologia - UFRN, Natal, RN, Brazil.}%

\date{\today}

\begin{abstract}
This article introduces nox-minima, a low-cost, three-dimensional paper dome that provides an alternative representation of the sky for astronomy education. Generated from precise astronomical data, the dome provides accurate local sky views for any date and location. Its assembly is simple, and the design is freely available at nox-minima.net/en~. Initial workshops with students and teachers confirm its effectiveness as a hands-on tool to explore the celestial sphere and cultural perspectives on the sky.\end{abstract}
\keywords{Nox-minima, planetarium, science communication.}

\maketitle


\section{Introduction}
Educational sky-watching activities are often accompanied by representations of the firmament, typically in the form of star charts such as that provided in references~\cite{Ford2025, Saraiva2011}, or through specialized applications like Stellarium~\cite{Stellarium2025}. Less frequently, the celestial sphere is represented as a globe. In principle, these representations are employed because they provide elements that support the understanding of the concepts involved in astronomical observation. However, they may also introduce conceptual difficulties inherent to their nature.

This paper introduces an alternative way of representing the sky: a small-scale celestial dome, named nox-minima. With its three-dimensional, nearly hemispherical shape, this representation preserves the relative spatial distances between celestial objects and avoids the distortions inherent in flat projections such as star charts. 
\newpage
By presenting the entire sky in miniature, the nox-minima dome offers a global view of the firmament, in contrast to the fragmented perspectives typical of mobile applications. It is designed so that celestial objects are viewed from an internal perspective, under the dome\footnote{After the development of the nox-minima dome, the websites~\cite{Upforfun2015,Krieger2023} were identified that propose similar representations of the night sky as paper domes designed to be observed from the inside.}. This topocentric viewpoint (alt-azimuth or local horizontal coordinate system) is pedagogically advantageous, as it directly corresponds to the three-dimensional perspective of an Earth-based observer. The zenith, for instance, is not at the center of a flat drawing but directly overhead. Such an advantage is absent in globe-based representations, which misleadingly suggest that the observer is located outside a sphere of stars. Highly immersive technological resources, such as digital planetariums and virtual reality environments, can also provide accurate and engaging experiences of the sky~\cite{Allen2024}; however, their high cost and limited availability restrict their use in many educational settings. The nox-minima dome, in turn, provides a tangible and accessible alternative.

The construction of the nox-minima dome relies on precise astronomical databases to generate, after geometric transformations, flat distorted representations of portions of the sky. Once assembled, these deformations result in celestial objects occupying coordinates (altitude and azimuth) very close to their correct positions for a chosen location and date. The technical details underlying the dome’s design will be briefly discussed in Section 2.

The dome’s tangible nature and spatial fidelity make it a valuable tool for realistic approaches, preparing students for direct sky observations or even replacing field activities when weather conditions are unfavorable. Another promising application is the use of the dome to represent skies from past epochs or to reproduce asterisms from indigenous cultures. These and other pedagogical possibilities enabled by the dome are presented in Section 3.

Assembling the dome is itself a valuable learning activity. It provides training in fine motor skills—important for the development of mathematical abstraction~\cite{Carbonneau2013} yet often overlooked in contemporary education. It also creates opportunities to explore concepts of plane and spatial geometry. Handling a miniature firmament stimulates curiosity and learning, as observed in workshops conducted with students at different educational levels. A brief report on these early workshops is presented in Section 4, followed by the concluding remarks in Section 5.

\section{The construction of the nox-minima dome}

With astronomical data at hand, and using a combination of mathematics and programming, it is possible to generate the images needed to print and assemble the stellar dome. This section presents some details of its construction.

\subsection{Astronomical database}

Several star catalogs have been compiled for astronomical purposes. One of the most important is the Hipparcos catalog~\cite{Perryman1997}, the result of a pioneering mission by the European Space Agency (ESA). This project provided precise measurements of position, parallax, and proper motion for about 120,000 stars, with an accuracy of 1 milliarcsecond. Since 1997, this database has been publicly available. More recent catalogs offer even greater precision and include a far larger number of stars and additional parameters.

The positions and magnitudes of cataloged stars, as well as Solar System bodies, must be calculated for a given observing site and moment in time. In this work, these computations are carried out using the Python library \texttt{SkyField}~\cite{Rhodes2019}. According to its documentation, the stellar positions generated by \texttt{SkyField} for the Hipparcos catalog are consistent with those of the Astronomical Almanac of the United States Naval Observatory (USNO), with discrepancies on the order of 0.5 milliarcsecond. For comparison, the angular diameter of Sirius as seen from Earth is 5.9 milliarcseconds. Because of this level of accuracy, the data from \texttt{SkyField} are suitable even for scientific research, and more than sufficient for educational purposes.

A complete sky map also requires additional elements, such as constellation lines and names, the ecliptic, celestial poles, and a depiction of the Moon with its phase and orientation. All of these can be obtained or derived using \texttt{SkyField}.

\subsection{Geometric transformations and projections}

Ideally, the dome would be assembled from flat paper segments, on which the symbols representing stars and auxiliary lines would be printed. However, because the sphere has non-zero Gaussian curvature, it cannot be unfolded into a flat surface~\cite{Henderson1996}. Dividing the sphere into segments does not solve this problem, since each segment itself also has curvature. In practice, these segments are printed on flat media and would require physical deformation to conform to the sphere. The elasticity of common bond paper, measured by Young’s modulus, is too low compared to other materials (such as vinyl), making such deformation unfeasible. In traditional globe construction, thinner types of paper are typically used, moistened during application to accommodate the necessary curvature.

To circumvent this limitation when using standard dry paper, the spherical surface can be approximated by a developable surface. In this approach, the hemispherical dome was partitioned into 72 regions (12 spherical triangles and 60 spherical quadrilaterals), and each region was replaced by a flat surface tangent to the sphere at the midpoint of the corresponding region, as illustrated in Figure~1. The number of regions was chosen to ensure minimum robustness of the object without making its assembly excessively laborious.

\begin{figure}[h]
\centering
\includegraphics[width=0.8\textwidth]{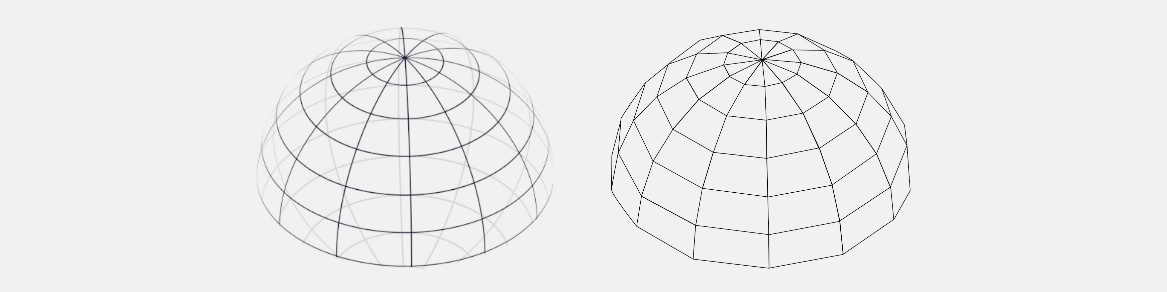}
\caption{On the left, the 72 regions used for the planar approximation of the hemispherical dome. On the right, the approximation using flat figures (60 quadrilaterals and 12 triangles). Image produced by the author.}
\end{figure}

The projection of a point $P$ on the sphere onto the dome is defined by the intersection of the tangent plane at the center of the region closest to $P$ with the extension of the radius passing through $P$, as illustrated in Figure~2. This is known as the gnomonic projection (or central projection), whose earliest use is attributed to Thales of Miletus~\cite{Snyder1987}. The gnomonic projection has the property of mapping great circles as straight lines and vice versa. Because of this property, it has found convenient applications in fields such as navigation and meteor trajectory analysis.

\begin{figure}[h]
\centering
\includegraphics[width=0.7\textwidth]{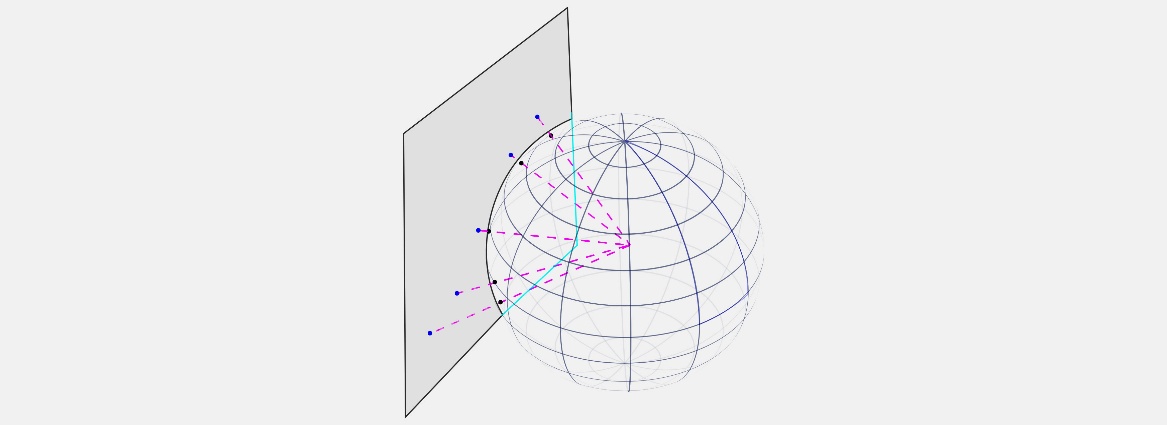}
\caption{Projection of celestial objects onto the tangent plane in each region. Image produced by the author.}
\end{figure}

In constructing the dome, 72 local gnomonic projections are carried out. Care is taken to exclude projections of points located in the opposite hemisphere of the considered region (corresponding to stars below the horizon). By symmetry, the intersection between two adjacent flat regions at the same altitude is the straight-line segment resulting from the projection of their shared spherical boundary, as shown in Figure~3. Consistently, this straight segment represents the arc of the great circle separating the neighboring regions.

\begin{figure}[h]
\centering
\includegraphics[width=0.7\textwidth]{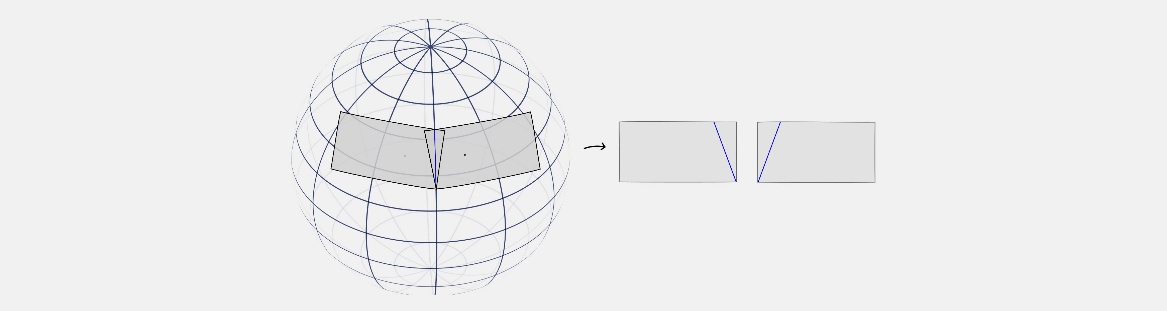}
\caption{The boundary between two adjacent regions is projected as the straight line defined by the intersection of their tangent planes. Image produced by the author.}
\end{figure}

After performing the projections, a sequence of transformations is applied in each tangent plane to obtain Cartesian coordinates $(x, y)$ on a single plane, corresponding to the sheet of paper, as suggested in Figure~4 (steps A to D). At the end of these transformations, the sphere is mapped into flat segments.

\begin{figure}[h]
\centering
\includegraphics[width=0.7\textwidth]{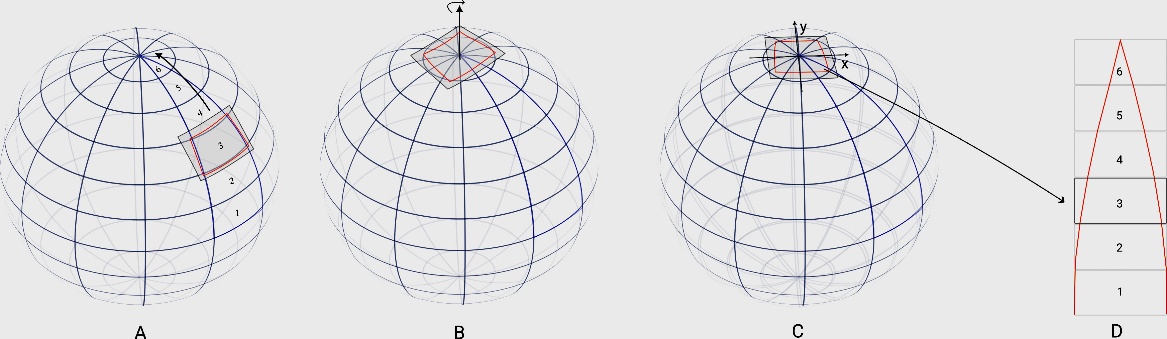}
\caption{Illustration of the transformations performed in each region of the sphere to produce the unfolded segments. Image produced by the author.}
\end{figure}

During the three-dimensional assembly, the procedure of aligning and gluing the straight edges of adjacent segments at the same altitude (Figure~5) ensures that the geometric constraints of neighboring planar regions are satisfied, thereby reconstructing the correct curvature of the dome.

\begin{figure}[h]
\centering
\includegraphics[width=0.7\textwidth]{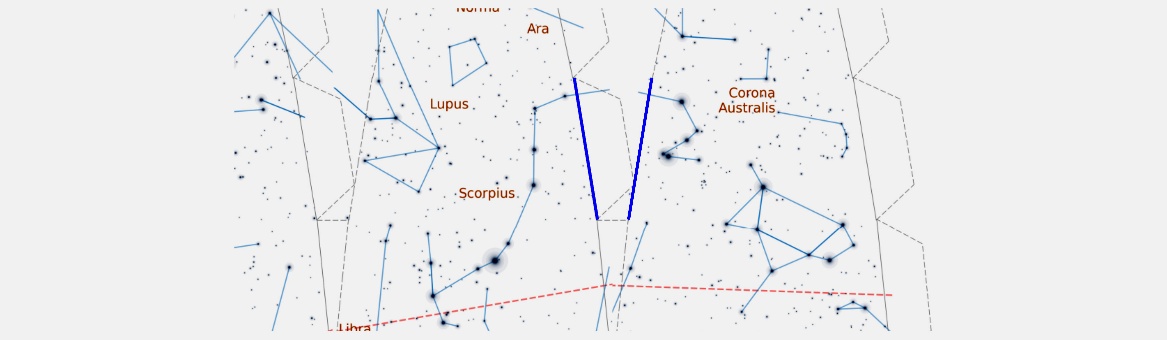}
\caption{By joining the straight-line edges (highlighted in bold blue) of adjacent regions at the same altitude, the dome is reconstructed with an approximately hemispherical shape. Image produced by the author.}
\end{figure}

\subsection{Generating dome figures: technical details}

The images for printing are generated using the Python library \texttt{matplotlib}. Cartesian coordinates of stars and planets are represented by circular symbols whose diameters vary according to the apparent magnitude of the respective object at the chosen location and time.  

Plotting constellation lines and auxiliary lines, such as the ecliptic, requires special treatment. When the endpoints of a line fall within the same segment (i.e., within one of the 12 regions of 30$^\circ$ azimuthal width), the line is drawn as a straight segment (the image of an arc of a great circle on the sky). If the endpoints fall in different segments, the angular interval must be divided into portions, one for each crossed segment. The endpoints within a given segment are then connected by a straight line.

Stars and planets located near the edges of segments are projected and represented in two or more adjacent pieces. This procedure prevents symbols from being split or truncated by the segment boundaries.  

The Moon requires special treatment for its phase, rotation, and illuminated orientation. From the phase angle provided by \texttt{SkyField}, one of 40 lunar images obtained from NASA~\cite{NASA2017} is selected. Variations due to lunar libration were not considered. The Moon’s rotation was approximated using the local inclination of the ecliptic, and the orientation of illumination (to one side or the other) was determined by analyzing the direction of minimum angular separation between the Moon and the Sun along the ecliptic.  

Finally, to simplify the dome’s assembly, the segments are rotated so that their lowest-altitude sectors (between 0$^\circ$ and 15$^\circ$) meet at the base, eliminating the need for additional cutting and gluing. The program outputs a PDF file with three pages, each containing four segments.

\begin{figure}[h]
\centering
\includegraphics[width=0.4\textwidth]{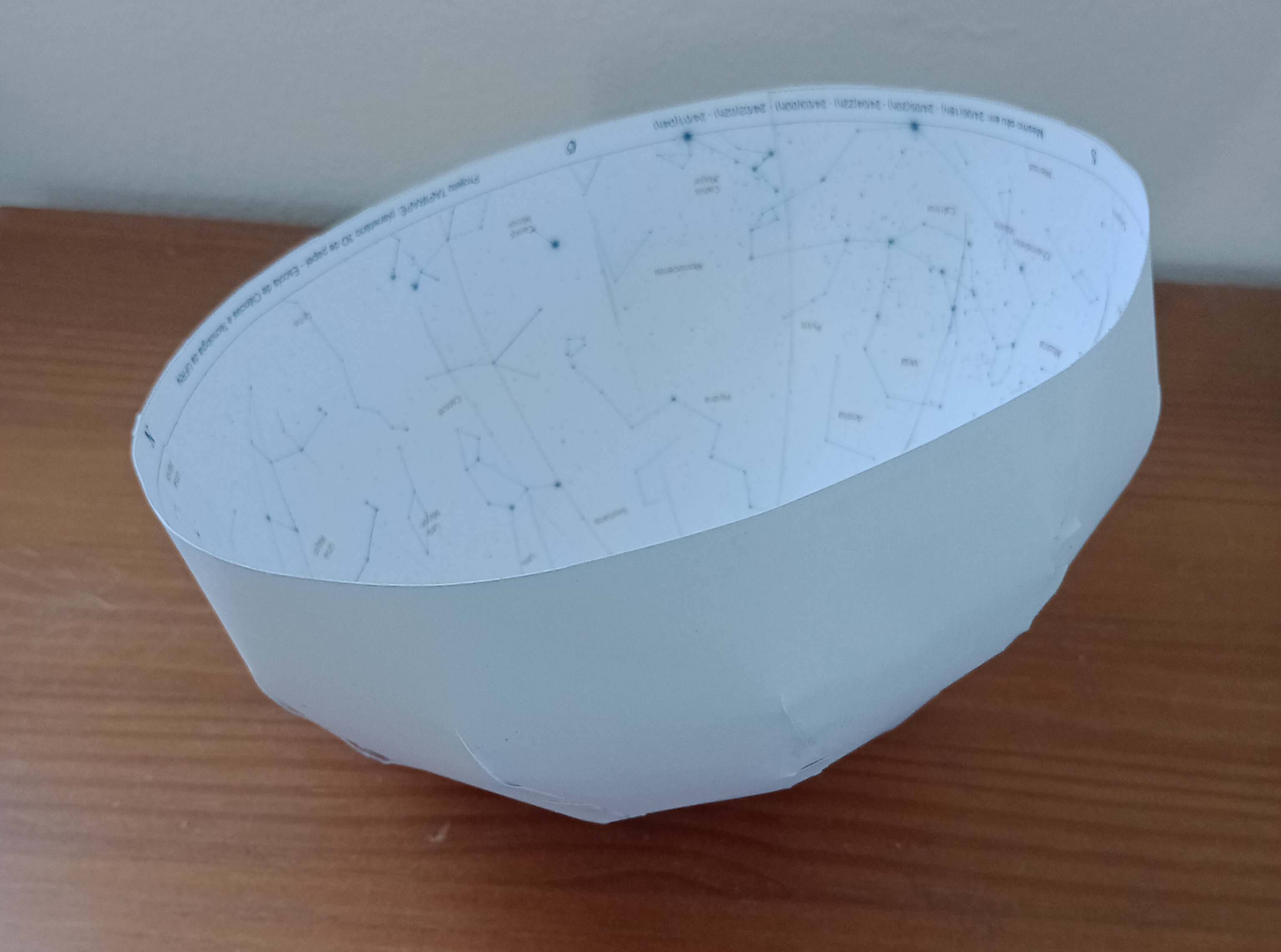}
\caption{The nox-minima dome. Source: author’s personal archive.}
\end{figure}

\subsection{Assembly of the dome}

The assembly process begins with choosing the appropriate paper for printing. To obtain a more durable dome, bond paper of 120 g/m$^2$ or, preferably, 180 g/m$^2$ is recommended. There is no need for special, more expensive paper types such as glossy or photographic paper. The current cost for assembling one dome (3 sheets of A4 or letter paper, 180 g/m$^2$) is about 0.15 USD. The maximum diameter of the dome printed on A4 paper is 24 cm.  

After printing, the pieces should be carefully cut along the dashed lines. White glue is applied to the flaps of each segment, spreading a thin layer with a finger to cover the entire flap surface. A useful tip is to glue all flaps of a segment before joining it to its neighbor. The assembly should always proceed from bottom to top, aligning constellation lines across adjacent segments without overlapping sky regions. With proper alignment, the paper naturally takes on the correct curvature. It is essential to remember that the stars face inward, not outward.  

Once each of the three groups of four segments is assembled, the groups are joined, again from bottom to top, preferably gluing all flaps of a segment before beginning the union. To avoid confusion in assembly order, the figure containing the western section has the letter “N” printed on its right side, indicating that it must be joined to the northern section. Groups are connected starting at the base, at right angles. With due care, the result is both aesthetically pleasing and structurally strong, with no gap at the zenith, as illustrated in Figure~6.  

\begin{figure}[h]
\centering
\includegraphics[width=0.5\textwidth]{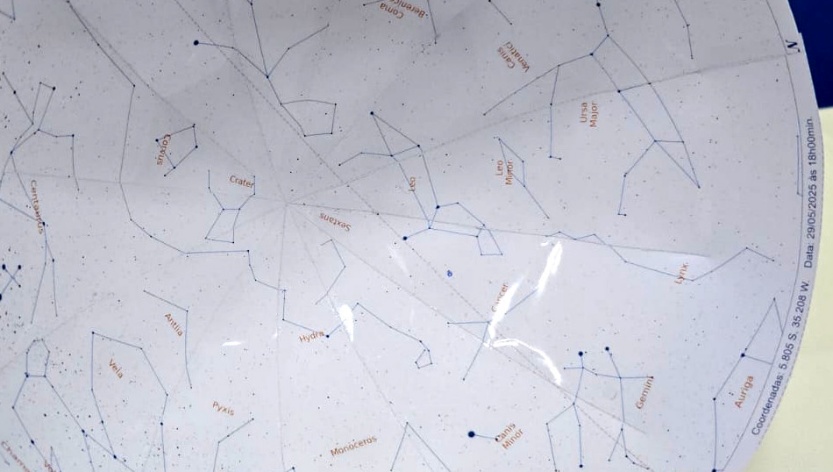}
\caption{Interior view of a nox-minima dome assembled during a workshop at the School of Science and Technology of the Federal University of Rio Grande do Norte, Brazil. Source: author’s personal archive.}
\end{figure}

\section{Pedagogical insights from the nox-minima dome}

From an educational perspective, the nox-minima dome can be explored in several ways. Below we outline approaches that are, in principle, facilitated or made possible by the dome’s spatial nature.

\subsection{Exploring spherical geometry}

The geometric aspects related to the sphere and the dome are of interest in their own right. Teachers may engage students in identifying the angles used in the dome’s construction and in relating them to angles observed in the sky (estimated, for example, using hand spans). The circumference of the dome can be estimated by measuring the straight-line heights of the flat regions that compose it. Concepts such as latitude and longitude, and their relation to standard spherical coordinates, can also be introduced.

The 72 flat regions of the dome provide a concrete (though approximate) representation of the directions of the $\theta$ and $\phi$ unit vectors, the basis vectors of the 72 tangent planes. An especially engaging challenge is estimating the area of each flat region. Teachers may also present the gnomonic projection, its properties, and applications at different levels of depth.

Since the Gaussian curvature of the sphere is non-zero at all points, it is impossible to unfold any spherical region into a plane. By contrast, cylinders and cones, though curved, can be unfolded completely. Why does this happen? Such a discussion can be tailored in depth according to the target audience.

\subsection{Experiencing the celestial sphere}

As noted in the introduction, the observational perspective of the dome corresponds to that experienced when viewing the night sky directly. In particular, the dome allows teachers to concretely illustrate the concept of the celestial sphere, emphasizing the varying distances between stars represented in adjacent regions.

Cardinal points and the zenith are easily mapped to their corresponding positions on the observer’s horizon. With directions established, teachers can introduce the brightest stars and most prominent constellations in each quadrant. Students gain a realistic sense of the ecliptic’s height above the horizon and zenith, as well as the points where it intersects the horizon. This information provides a basis for discussions on the transit times of the Sun, Moon, and planets across the sky.

Another important feature is the position of the celestial pole and its relationship to the observer’s latitude. For those in the Southern Hemisphere, strategies to locate the pole can be practiced and tested, such as extending the longer axis of the Southern Cross. At what angular distance does this extension pass from the pole? The answer can be visually estimated using the dome.

Teachers can also highlight the relative positions of constellations in the sky and introduce star-hopping techniques. Students may be given a “treasure map” to locate specific sky regions or objects, starting from easily recognizable points and working step by step toward more elusive targets.

\subsection{Recording observations on the dome}

The paper dome serves as a practical medium for annotations, making it a useful tool for recording observations. Naturally, the sky at a given hour will differ slightly from the sky an hour later or on the following night at the same time. Recognizing these changes is itself an important learning experience. For many activities, the same dome can be used for several hours in a night or across consecutive days at the same time of day.

Students can, for example, record the movement of the Moon and planets: Is the Moon approaching or moving away from the ecliptic? Does this relate to the occurrence of eclipses? Do planets always move in the same direction? Such questions can stimulate investigation and discussion.

Other possibilities include coloring stars and planets according to visual perception, marking the positions of predicted events (e.g., meteor showers), or building a log of observed phenomena (shooting stars, variable stars, International Space Station transits, or objects viewed through a telescope).

\subsection{Inclusive astronomy: touching the sky}

The dome’s tangible format also makes it adaptable for tactile perception by blind or visually impaired students. When printed on A3 paper, the dome’s maximum diameter increases to 34 cm, which facilitates recognition of smaller structures. Stars and Solar System bodies can be represented in relief using beads, grains of sand, or raised-ink pens. Horizon lines, constellation lines, and the ecliptic can be marked with threads of suitable thickness. The expectation is that the three-dimensional dome will enable recognition not only of individual celestial objects but also of their spatial orientation.

\subsection{Reproducing skies from different epochs and cultures}

Concrete reproduction of the sky from past epochs is particularly valuable for research presentations in fields such as archaeoastronomy. The main long-term cause of change in the sky is the shifting position of the poles due to Earth’s axial precession. Over a period of roughly 13,000 years, the north celestial pole drifts from its present vicinity near Polaris to a region near Vega in the constellation Lyra. In addition, the relative positions of stars gradually shift due to proper motion. These corrections can be computed using \texttt{SkyField}. Planetary and lunar positions are available as far back as 3000~BC, encompassing the skies of many ancient civilizations.

Another field that benefits from such concrete representations is ethnoastronomy, an interdisciplinary area that studies celestial observation systems of ancient or contemporary indigenous peoples \cite{Lima2014}. Asterisms represented as connecting lines between stars or sky regions can easily be incorporated into the nox-minima dome. At present, four asterisms from Brazilian indigenous traditions are available: the Rhea, the Old Man, the Northern Tapir, and the Deer~\cite{Afonso2005}. Figure~8 illustrates two of these constellations.

\begin{figure}[h]
\centering
\includegraphics[width=0.8\textwidth]{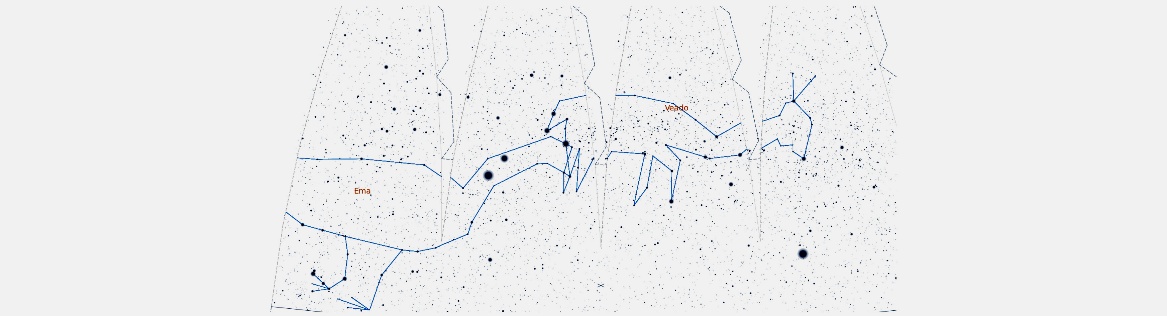}
\caption{Representation of the Rhea and the Deer, indigenous Brazilian asterisms. Image produced by the author.}
\end{figure}

New asterisms resulting from ongoing research can be incorporated. Bringing the symbolism of the southern sky, as conceived by Brazil’s native peoples, into the classroom is particularly valuable for students.

\section{First workshop reports with the nox-minima dome}

The website maintained by the author provides open access to images of the sky for any chosen location and date. Since its launch in July 2024, more than 800 images have been generated for dome assembly. During this period, the site has been accessed over 1,500 times from 35 countries. Approximately 50\% of visits come from Brazil. The United States (15\%), Germany (10\%), France (7\%), the United Kingdom (3\%), and Spain (2\%) are the main sources of international access. Interest has been noted from students, teachers, researchers across various fields, as well as from science museum administrators.

In Brazil, the first workshops with the nox-minima dome were conducted with students and teachers through the Universidade Estadual de Feira de Santana (UEFS), in Bahia, coordinated by researcher Marildo Pereira. Reports from these workshops, carried out with teachers in the Professional Master’s Program in Astronomy at UEFS, highlight the rich opportunities for conceptual discussion provided by the dome’s concrete representation of the sky. 

\begin{figure}[h]
\centering
\includegraphics[width=0.6\textwidth]{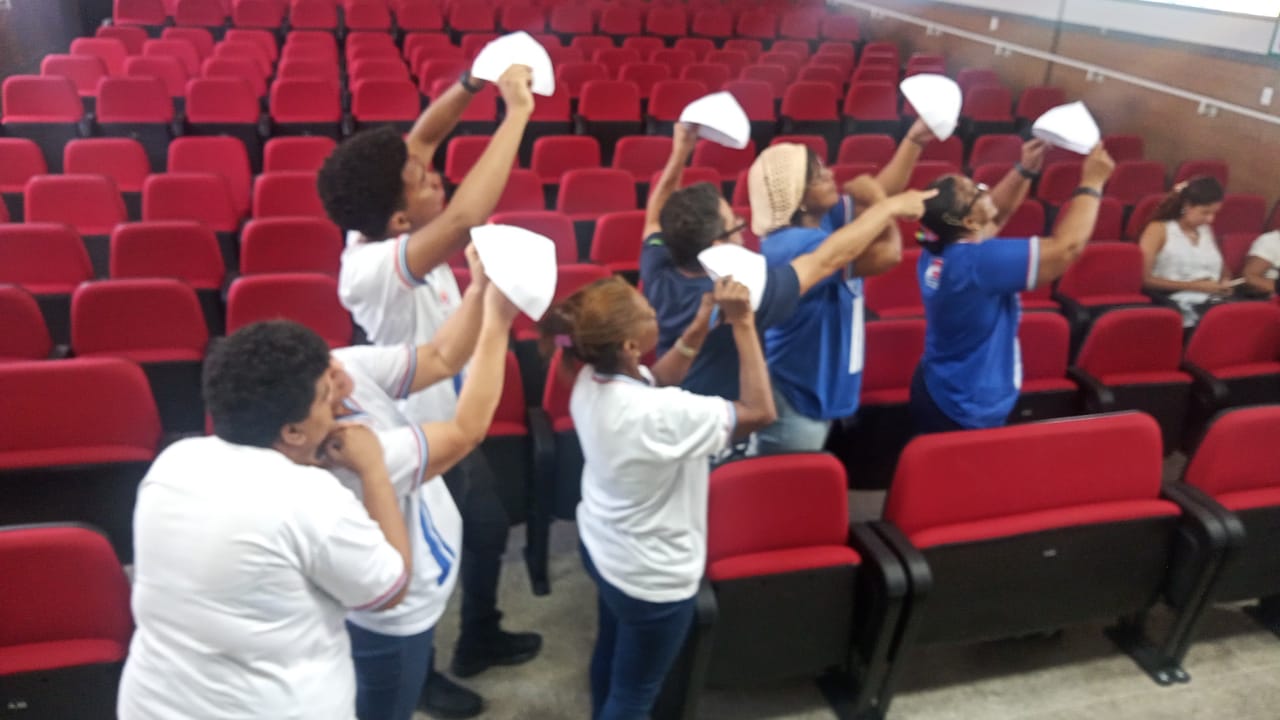}
\caption{Use of the nox-minima dome with students and teachers at the Colégio Estadual de Tempo Integral Professora Carminda Mascarenhas Vieira, Feira de Santana, Bahia. Source: author’s personal archive.}
\end{figure}

At the Federal University of Rio Grande do Norte (UFRN), workshops were conducted within the extension project \textit{Cometa Nordestino}, coordinated by researcher Leonardo Almeida. Different strategies were tested. During a visit to Escola Estadual Stoessel de Brito, in Maxaranguape, RN, domes were pre-assembled and distributed to high-school students. The one-hour workshop focused on recognizing the sky and presenting stories from Greek and indigenous mythology related to prominent sky regions on the day of the visit.  

In events held at UFRN’s School of Science and Technology, open to the general public, participants were invited to assemble the dome during the 90-minute workshop. The assembly instructions themselves provided an opportunity to address conceptual elements. Students engaged with the activity in a relaxed yet focused manner, aiming for a satisfactory final result. Based on these experiences, it is recommended that one hour be reserved for dome assembly in workshops with young people or adults, even when working in pairs. Teachers should monitor inevitable comparisons between domes, assisting groups that encounter greater difficulties. In the end, all participants expressed pride in their completed dome and interest in the information it contained.

\section{Concluding remarks}

This paper presented the nox-minima dome, addressing the technical, practical, and educational aspects of its construction. The dome’s accuracy results from its geometric design and the reliability of the astronomical data employed. Its low cost and carefully planned geometry make it feasible for classroom use within the Brazilian educational context. Together, these features yield a learning tool with considerable didactic potential.  

The complete process—from generating the images, to assembling the dome, to employing it in the classroom—has been validated through the first workshops carried out in schools and universities. The strong interest from both domestic and international audiences, along with the positive feedback from workshop participants, confirms the dome’s potential as a resource for astronomy education.  

New functionalities and display options will continue to be developed based on the current code, which will eventually be made publicly available through an open repository.

\end{document}